\documentclass[11pt]{JHEP3}

\JHEPspecialurl{http://jhep.sissa.it/JOURNAL/JHEP3.tar.gz}

\usepackage{epsfig,multicol}
\usepackage{amsmath,amssymb}
\usepackage{mathrsfs}

\newcommand\fverb{\setbox\fverbbox=\hbox\bgroup\verb}
\newcommand\fverbdo{\egroup\medskip\noindent%
			\fbox{\unhbox\fverbbox}\ }
\newcommand\fverbit{\egroup\item[\fbox{\unhbox\fverbbox}]}
\newbox\fverbbox

\newcommand{\pslash}{p\kern-1ex /}
\newcommand{\qslash}{q\kern-1ex /}
\newcommand{\lslash}{l\kern-1ex /}
\newcommand{\sslash}{s\kern-1ex /}
\newcommand{\kaslash}{k_a\kern-2ex /}
\newcommand{\kbslash}{k_b\kern-2ex /}
\newcommand{\Dslash}{\mathcal{D}\kern-1.5ex /}

\newcommand{\beqa}{\begin{eqnarray}}
\newcommand{\eeqa}{\end{eqnarray}}

\newcommand{\ba}{\begin{eqnarray}}
\newcommand{\ea}{\end{eqnarray}}
\newcommand{\be}{\begin{equation}}

\title{5-loop Konishi from linearized TBA and the XXX magnet }

\author{J\'anos Balog and \'Arp\'ad Heged\H us\\
Research Institute for Particle and Nuclear Physics, 
Hungarian Academy of Sciences,
H-1525 Budapest 114, P.O.B. 49, Hungary\\}

\received{\today} 		
\accepted{\today}	

\abstract{
Using the linearized TBA equations recently obtained in arXiv:1002.1711
we show analytically that the 5-loop anomalous dimension of the Konishi
operator agrees with the result obtained previously from the generalized 
L\"uscher formulae.
The proof is based on the relation between this linear system and the XXX 
model TBA equations.
}

\begin{document}

\section{Introduction}

One of the most important problems in testing the AdS/CFT correspondence 
\cite{adscft} is to understand
the finite size spectrum of the $AdS_5 \times S^5$ superstring. For large 
volumes the asymptotic
Bethe Ansatz (ABA) describes the spectrum of the model \cite{BS}. 
It takes into account all power like
corrections in the size, but neglects the exponentially small wrapping corrections \cite{AJK}.

In \cite{BJ08} it was shown that the leading order wrapping corrections can also be expressed by the infinite
volume scattering data through the generalized L\"uscher formulae \cite{Luscher85}.
In \cite{BJ08} the 4-loop anomalous dimension of the Konishi operator was obtained by means of the generalized
L\"uscher formulae in perfect agreement with direct field theoretic computations \cite{Sieg,Vel}.
Subsequently wrapping interactions computed from L\"uscher corrections were found to be crucial for the
agreement of some structural properties of twist two operators \cite{Bajnok:2008qj} with LO and NLO BFKL 
expectations \cite{KL02,40_5}.

More recently \cite{BJ09} the 5-loop wrapping correction to the anomalous dimension of the Konishi operator was 
also computed from the generalized L\"uscher approach yielding the result:
\begin{equation*}
\Delta^{(10)} = \Delta_{\rm asympt}^{(10)}
+ g^{10} \Big\{
-\frac{81 \zeta(3)^2}{16}+\frac{81 \zeta(3)}{32}-\frac{45 \zeta(5)}{4}+\frac{945 \zeta(7)}{32}-\frac{2835}{256} 
\Big\},
\end{equation*}
with $g$ being the coupling constant related to the 't Hooft coupling $\lambda$ through $\lambda=4\pi^2g^2$.
Later the 5-loop result has been extended to the class of twist two operators as well
\cite{Lukowski:2009ce}. After analytic continuation to negative values of the spin this
gave nontrivial agreement with the predictions of the BFKL equations \cite{KL02}.

Due to the integrability of the string worldsheet theory, the Thermodynamic Bethe Ansatz (TBA) approach for the 
mirror model \cite{AJK,AF07} offers a tool to investigate the spectrum of the string theory. The TBA 
equations were derived first for the ground state \cite{AF09a,AF09b,AF09d,Bombardelli:2009ns,GKKV09}. Later 
using an analytic continuation trick \cite{DT} they were 
extended to excited states lying in the $sl(2)$ sector of the theory \cite{GKKV09,Arutyunov:2009ax} as well.
The TBA equations passed some tests both in the weak and in the strong coupling limit.
In the strong coupling limit it was shown \cite{Gromov} that the TBA equations reproduce correctly the 1-loop
string energies in the quasi-classical limit. 
In the weak coupling regime they  
give (by construction) the same leading order wrapping corrections in $g$ as predicted by the generalized 
L\"uscher 
formulae.

However, to extract the next to leading order wrapping correction in $g$ from TBA is more difficult
as at this order the modification of the ABA equations must be taken into account. 
In the TBA approach the modified ABA equations depend also on the asymptotically non-vanishing $Y$-functions
(which satisfy non-trivial coupled equations even in the small $g$ limit), making the next to leading
order calculation of wrapping interactions a non-trivial task.

In the TBA formulation of the finite size problem the energy of an $N$-particle state takes the form:
\begin{equation} \label{E}
E=J+\sum_{i=1}^N{\cal
E}(p_i)-\frac{1}{2\pi}\sum_{Q=1}^{\infty}\int_{-\infty}^{\infty}{\rm
d }u\frac{d\widetilde{p}^Q}{du}\log(1+Y_Q)\,,  
\end{equation} 
where $J$ is
the angular momentum carried by the string rotating around the
equator of $S^5$, $\widetilde{p}^Q$ is the mirror momentum
and the functions $Y_Q$ are the unknown functions (Y-functions) associated to the mirror $Q$-particles,
futhermore 
\begin{equation} {\cal E}(p)=\sqrt{1+4g^2\sin^2\frac{p}{2}}
\end{equation}
is the dispersion relation of the string theory particles.

In this paper we will focus on the $g^{10}$ order computation of the anomalous dimension (energy)
of the Konishi operator $\mbox{Tr}(D^2 Z^2\!-\!(DZ)^2)$ and expanding the considerations of \cite{AFS}
we prove analitically that the TBA equations and the generalized L\"usher formulae 
of \cite{BJ09} give the same result for the 5-loop anomalous dimension of the Konishi operator.
This operator corresponds to the $N=J=2$ choice in (\ref{E}). For its TBA equations see 
\cite{GKV09b,Arutyunov:2009ax}.
In this paper we will use the notations and conventions used in \cite{Arutyunov:2009ax}.
  
It is known for the Konishi operator that in the weak coupling regime the wrapping corrections start at the 
order of $g^8$ thus the ABA equations for the momenta get corrections from 
wrapping at $g^8$ order, i.e. 
$\delta p_k \sim g^8$, where $\delta p_k$ is the wrapping correction to the asymptotic value of the $k$th 
momentum.

The energy formula (\ref{E}) can be expanded around the asymptotic solution if the $Y$-functions are small.
This happens either for large $J$ as the $Y_Q$-functions are exponentially small in this limit or at fixed
$J$ for small $g$. From the asymptotic solution of the TBA equations \cite{GKV09} it is known that $Y_Q \sim 
g^8$,
this is why up to $O(g^{10})$ it is enough to take into account only the first term, linear in $Y_Q$, in the 
series 
expansion of the integral term of (\ref{E}).
\begin{equation} \label{dE}
E \simeq J+\sum_{i=1}^2{\cal E}(p^{ABA}_i)+ \sum_{i=1}^2 \frac{\partial {\cal E}}{\partial p_i} \bigg|_{ABA} 
\! \! \!  \delta  p_i -\frac{1}{2\pi}\sum_{Q=1}^{\infty}\int_{-\infty}^{\infty}{\rm
d }u\frac{d\widetilde{p}^Q}{du} \, Y_Q \,+O(g^{12}),  
\end{equation}  
where $p_i^{ABA}$ denotes the solution of the ABA and the derivative of ${\cal E}$ must be taken at the 
asymptotic values of the momenta.
As  the function $\frac{d\widetilde{p}^Q}{du}$ starts at $O(1)$ in $g^2$ and $\frac{\partial {\cal E}}{\partial 
p_i} \bigg|_{ABA} \sim g^2$ it can be seen that only the asymptotic form of the $Y_Q$ functions contribute 
to the wrapping correction in leading order and the momentum perturbations start to play a role only at 
$O(g^{10})$.
 
Taking the asymptotic form of the $Y_Q$ functions given in \cite{GKV09} it is easy to see that all terms in the
above energy expression identically
agree with those of ref. \cite{BJ09} except the one containing the momentum 
correction. Thus only the momentum quantization equations should be compared to see whether both approaches 
give the same result for $\delta p_i$.

In a recent publication \cite{AFS} this agreement was verified by numerically solving the linearized TBA 
equations.
We will use the results (and notations) of this paper.

Let $u_k=u_k^o+\delta u_k$, where $u_k^o$ is the asymptotic value of the $u_k$ and $\delta u_k \sim g^8$
is its wrapping correction. Then $\delta u_k$ satisfies the equation: 
$$
\sum_{j=1}^2 \frac{\delta {\rm ABA}(u_k,\{u_i\})}{\delta u_j} \bigg|_{u_i=u_i^o} \, \delta
u_j+\Phi_k^{(8)}=0\, 
$$
where $\Phi_k^{(8)}$ is the $O(g^8)$ correction to the ABA. For small $g$ all $Y_Q$ functions
are small and the TBA equations can be linearized around the asymptotic solution.

In \cite{AFS} it has been shown that at $O(g^8)$ the linear problem for the functions
associated to the $vw$-strings decouples from the other type of variables and takes the form 
\begin{equation}
\label{LYvw} 
{\mathscr Y}_{M|vw}=  (A_{M-1|vw}{\mathscr Y}_{M-1|vw}+ A_{M+1|vw}{\mathscr Y}_{M+1|vw})\star s
-Y_{M+1}^o\star s,
\ \ \ \ {\mathscr Y}_{0|vw}=0,
\ \ \ \ M=1,2,\dots,
\end{equation}
where ${\mathscr Y}_{M|vw}$ is the $O(g^8)$ perturbation of the asymptotic $Y_{M|vw}^o$ function defined by 
the formula $Y_{M|vw}=Y_{M|vw}^o \, (1+{\mathscr Y}_{M|vw})$, 
$A_{M|vw}=\lim\limits_{g \to 0} \frac{Y_{M|vw}^o}{1+Y_{M|vw}^o}$ and it is given explicitly by 
{\footnote{We simplified the formula (2.15) of ref.\cite{AFS} to make its
relation to the XXX magnet apparent.}}
\begin{equation} \label{A_{M|vw}}
A_{M|vw}(u)=\frac{M(M+2)}{(M+1)^2}\,\,
\frac{(u^2-w^2+M^2-1)\,(u^2-w^2+(M+2)^2-1)}
{[u^2-w^2+(M+1)^2]^2+4w^2-4M(M+2)},
\end{equation}
where 
\begin{equation}
u_1=-u_2=w=\frac{1}{\sqrt{3}} 
\label{spec}
\end{equation}
is the $O(1)$ solution of the ABA for the Konishi state.  
Furthermore $\star$ denotes convolution, $s$ is the TBA kernel $s(x)=\frac{1}{4 \, \cosh(\frac{\pi}{2}x)}$, and 
finally in the source term of the linear problem $Y_Q^o$ is the leading, $O(g^8)$, asymptotic expression
of the $Y_Q$ functions:
{\begin{eqnarray} \label{YQg8}
&& Y_Q^o(u)=g^8 \,
\frac{64 \, Q^2[-1+Q^2+u^2-w^2]^2}{(Q^2+u^2)^4[(Q-1)^2+(u-w)^2][(Q+1)^2+(u-w)^2]}\times
\\ \nonumber
&& \hspace{5cm}\times
\frac{1}{[(Q-1)^2+(u+w)^2][(Q+1)^2+(u+w)^2]}. \end{eqnarray}} \hspace{-0.255cm} \noindent

It turns out \cite{AFS}
that apart from the $Y_Q^o$ functions and ${\mathscr Y}_{1|vw}$ 
no perturbations of the other $Y$-functions enter the final formula for 
$\delta {\cal R}_k=-\Phi_k^{(8)}$:
\begin{equation} \label{TBAresult}
\begin{split}
\delta{\cal R}_k&=\frac{1}{\pi}\sum_{m=1}^\infty\,\int_{-\infty}^\infty{\rm d}
u\,Y^o_m(u)\,\frac{u-u_k}{(m+1)^2+(u-u_k)^2}+\rho_k\\
&+\frac{1}{\pi}\sum_{m=1}^\infty\,\int_{-\infty}^\infty{\rm d}
u\,Y^o_{m+1}(u)\,\left\{{\cal F}_m(u-u_k)-
\frac{u-u_k}{m^2+(u-u_k)^2}\right\},
\end{split}
\end{equation}
where
\begin{equation}
{\cal F}_m(u)=\frac{-i}{4}\left\{\psi\left(\frac{m+iu}{4}\right)
-\psi\left(\frac{m-iu}{4}\right)-\psi\left(\frac{m+2+iu}{4}\right)
+\psi\left(\frac{m+2-iu}{4}\right)\right\}
\end{equation}
with the usual $\psi$ function $\psi(z)=\Gamma^\prime(z)/\Gamma(z)$ and $\rho_k$ is the contribution
coming from the $Y_{M|vw}$-functions:
\begin{equation}
\rho_k=\int_{-\infty}^{\infty} {\rm d}u\,
A_{1|vw}(u){\mathscr Y}_{1|vw}(u)\,\frac{1}{2\sinh\frac{\pi}{2}(u-u_k)}.
\label{rho0}
\end{equation}
On the other hand the generalized L\"uscher approach provides \cite{BJ09} the following expression for 
$\Phi_k^{(8)}$:
\begin{eqnarray} \label{Luscher}
&&\Phi^{(8)}(u_k)=\sum_{M=1}^{\infty}\int_{-\infty}^{\infty}{\rm d}u\,Y_M^{o}(u)\times\\
\nonumber &&~~~~~~\times \frac{1}{\pi}\Big[
-\frac{u-u_k}{(M+1)^2+(u-u_k)^2}-\frac{u-u_k}{(M-1)^2+(u-u_k)^2}+\frac{u_k}{-1+M^2+u^2-u_k^2}\Big]\,
. \end{eqnarray}
In \cite{AFS} it has been numerically verified that $\Phi_k^{(8)}$ given by (\ref{TBAresult}) and 
(\ref{Luscher})
agrees. In this paper we will show this fact analytically. 
The key point of the proof is to recognize that the coefficient functions $A_{M|vw}$ of the linear problem
(\ref{LYvw}) are related to the $Y$-functions of the inhomogeneous spin-$\frac12$ XXX chain \cite{Wieg} and
that (with a different source term) the linear problem (\ref{LYvw}) is identical to the
variation of the TBA equations{\footnote{In this case the term TBA is used in the sense of finite size 
effects.}}
of the XXX magnet with respect to the inhomogeneity parameters.
Exploiting these facts we can express the quantity $\Phi_k^{(8)}$ by the $Y$-functions of the XXX magnet and
show that the formulae (\ref{TBAresult}) and (\ref{Luscher}) are identical (up to a sign).

\section{Linearized TBA equations}

Let us rewrite the linearized AdS TBA system (\ref{LYvw}) as follows:
\begin{equation}
D_m\,\delta L_m-s\star(\delta L_{m+1}+\delta L_{m-1})=-s\star Y^o_{m+1},
\qquad m=1,2,\dots
\label{lin}
\end{equation}
Note that our unknown functions $\delta L_m(u)$ are rescaled (by $A_{m|vw}(u)$) with respect to the
ones used in ref. \cite{AFS} and the coefficient functions $D_m(u)$ are
the inverses of the functions $A_{m|vw}(u)$ given by (\ref{A_{M|vw}}).
In this note we will only use the fact that the $Y^o_m(u)$ functions are regular and even in $u$,
but their explicit form (\ref{YQg8}) is not needed. In (\ref{lin}) $\delta L_0=0$ by
convention and we also note that $\delta L_1(u_k)=0$ because of the rescaling by $A_{1|vw}(u)$,
since the latter function vanishes at $u=u_k$. We first have to solve 
(\ref{lin}) and then the relevant quantity to be calculated is
\begin{equation}
\rho_k=\int_{-\infty}^\infty{\rm d}u\,\,\frac{\delta L_1(u)}
{2\sinh\frac{\pi}{2}(u-u_k)}.
\label{intsol}
\end{equation}
No principal value prescription is needed since the integrand is regular
at $u=u_k$. If we can calculate $\rho_k$ then the leading correction to
the Bethe-Yang quantization is given by (\ref{TBAresult}).

To avoid the singularities coming from $D_1(u)$ at $u_k$ 
 we shift the integration contour in the imaginary direction by a small amount $i\gamma$:
\begin{equation}
D^\gamma_m\,\delta L^\gamma_m-
s\star(\delta L^\gamma_{m+1}+\delta L^\gamma_{m-1})=-s^\gamma\star Y^o_{m+1},
\qquad m=1,2,\dots
\label{lin2}
\end{equation}
Here we use the notation $f^\gamma(u)=f(u+i\gamma)$ for any function $f(u)$. 
Although we need to solve (\ref{lin2}) in a particular case only, it turns out
to be useful to study the corresponding general linear problem, for a general
(infinite) vector of unknowns $\xi$ and arbitrary (infinite) source vector
$j$:
\begin{equation}
{\bf M}\,\xi=j,
\label{eq}
\end{equation}
where the operator matrix is given by
\begin{equation}
{\bf M}=\begin{pmatrix}
D^\gamma_1&-s\star&0&\dots\\
-s\star&D^\gamma_2&-s\star&\dots\\
0&-s\star&D^\gamma_3&\dots\\
{}&\vdots&{}&{}
\end{pmatrix}.
\label{opmatrix}
\end{equation}
In our case the unknowns are
\begin{equation}
\xi=
\begin{pmatrix}
\delta L^\gamma_1\\
\delta L^\gamma_2\\
\vdots
\end{pmatrix}
\end{equation}
and the source term is of the form
\begin{equation}
j=I=
\begin{pmatrix}
-s^\gamma\star Y^o_2\\
-s^\gamma\star Y^o_3\\
\vdots
\end{pmatrix}.
\label{source1}
\end{equation} 
The operator matrix ${\bf M}$ is symmetric, ${\bf M}^T={\bf M}$. Therefore, 
assuming that the inverse of ${\bf M}$ exists uniquely{\footnote{See 
Appendix A about the existence and unicity of the inverse of ${\bf M}$.}} 
we can formally solve (\ref{eq}) as
\begin{equation}
\xi={\bf R}\,j,\qquad\quad{\bf R}={\bf M}^{-1},
\label{sol}
\end{equation}
such that the inverse operator ${\bf R}$ is also symmetric: ${\bf R}^T={\bf R}$.
Writing the solution (\ref{sol}) in components we have
\begin{equation}
\xi_m(x)=\sum_{m^\prime=1}^\infty\,\int_{-\infty}^\infty{\rm d}y\,
R_{mm^\prime}(x,y)\,j_{m^\prime}(y),
\label{sol1}
\end{equation}
where, due to the symmetry of the operator, the kernels satisfy
\begin{equation}
R_{mm^\prime}(x,y)=R_{m^\prime m}(y,x).
\label{sym}
\end{equation}
Using this notation, we have
\begin{equation}
\begin{split}
\rho_k=\int_{-\infty}^\infty{\rm d}u\,&\frac{\delta L_1^\gamma(u)}
{2\sinh\frac{\pi}{2}(u+i\gamma-u_k)}\\
&=\sum_{m=1}^\infty\,
\int_{-\infty}^\infty{\rm d}u\,
\int_{-\infty}^\infty{\rm d}v\,
\frac{1}{2\sinh\frac{\pi}{2}(u+i\gamma-u_k)}R_{1m}(u,v)I_m(v).
\end{split}
\label{rho}
\end{equation}
In this paper we will compute the relevant quantity $\rho_k$ given by (\ref{rho})
without solving explicitly the linearized TBA equations (\ref{lin}).
This can be done by recognizing that an explicitly solvable auxiliary
linear problem can be defined via the XXX model which is of the form (\ref{eq})
with a special right hand side $j$. 
This linear problem
is the linearization of the TBA system corresponding to the XXX model
such that the coefficient functions $D_m$ are related to the XXX model
Y-functions. The construction and the solution of this linear problem is
given in the next section.

\section{XXX model TBA equations}

The XXX model transfer matrix eigenvalue relevant for our considerations is
\begin{equation}
t_m(u)=(m+1)\left\{(u-u_1)(u-u_2)+m(m+2)\right\}, \qquad m=-1,0,1,2,\dots.
\end{equation}
This is a zero isospin solution of the T-system equations for the inhomogeneous
XXX spin chain{\footnote{Here we consider the case when the inhomogeneities $u_1$ and $u_2$ are real.}} 
of length 2 (the corresponding Baxter Q-operator has one real 
Bethe root):
\begin{equation}
t_m(u+i)\,t_m(u-i)=t_{m+1}(u)\,t_{m-1}(u)+t_0(u+(m+1)i)\,t_0(u-(m+1)i), \qquad m=0,1,2,\dots
\end{equation}
The Y-system elements are given by the usual definitions
\begin{equation}
y_m(u)=\frac{t_{m+1}(u)\,t_{m-1}(u)}{t_0(u+(m+1)i)\,t_0(u-(m+1)i)},
\end{equation}
\begin{equation}
1+y_m(u)=\frac{t_m(u+i)\,t_m(u-i)}{t_0(u+(m+1)i)\,t_0(u-(m+1)i)}
\end{equation}
and satisfy the Y-system equations
\begin{equation}
y_m(u+i)\,y_m(u-i)=[1+y_{m+1}(u)]\,[1+y_{m-1}(u)].
\end{equation}

Now the crucial observation is that with this solution 
\begin{equation}
D_m(u)=\frac{1}{A_{m|vw}(u)}=\frac{1+y_m(u)}{y_m(u)},
\label{1perA}
\end{equation}
where the functions $A_{m|vw}(u)$ are given by (\ref{A_{M|vw}}).
More precisely, (\ref{1perA}) holds for the symmetric case (\ref{spec}).

Our T-functions (except $t_0$) have no physical roots 
(zeroes with imaginary parts less than unity) if
\begin{equation}
\left\vert\frac{u_1-u_2}{2}\right\vert<\sqrt{2}
\end{equation}
and therefore (for $m\geq1$) only $y_1(u)$ has physical roots.
The corresponding TBA equations are of the form
\begin{equation}
y_m(u)=\left\{t(u-u_1)\,t(u-u_2)\right\}^{\delta_{m1}}\,
\exp\left\{s\star(L_{m+1}+L_{m-1})(u)\right\},
\label{TBA}
\end{equation}
where
\begin{equation}
t(u)=\tanh\frac{\pi}{4}u\qquad{\rm and}\qquad L_m(u)=\ln(1+y_m(u)).
\end{equation}
Taking the derivative ($\partial_k$) of (\ref{TBA}) with respect to 
$u_k$ gives 
\begin{equation}
\frac{\partial_ky_m(u)}{y_m(u)}=-\frac{\pi\,\delta_{m1}}
{2\sinh\frac{\pi}{2}(u-u_k)}
+\left(s\star\partial_kL_{m+1}\right)(u)
+\left(s\star\partial_kL_{m-1}\right)(u).
\label{der}
\end{equation}
After shifting the $u$ variable by $i\gamma$ and making 
the specialization{\footnote{In the rest of the paper $\partial_k$ is 
understood as first taking
the derivative with respect to $u_k$ and then taking the specialization
corresponding to (\ref{spec}).}} 
(\ref{spec}) we get the auxiliary linear problem which is precisely of the
form (\ref{eq}) with
\begin{equation} \label{3.11}
\xi_m(u)=\partial_kL^\gamma_m(u)
\end{equation}
and
\begin{equation} \label{3.12}
j_m(u)=-\frac{\pi\,\delta_{m1}}{2\sinh\frac{\pi}{2}(u+i\gamma-u_k)}.
\end{equation}
Substituting (\ref{3.11}) and (\ref{3.12}) into (\ref{sol1}) we get  a relation
between the solution (\ref{3.11}) and certain matrix elements of the inverse operator
\begin{equation}
\partial_kL^\gamma_m(u)=-\frac{\pi}{2}\int_{-\infty}^\infty{\rm d}v\,
\frac{R_{m1}(u,v)}{\sinh\frac{\pi}{2}(v+i\gamma-u_k)}.
\label{Rm1}
\end{equation}

\section{Calculation of $\rho_k$}


From (\ref{rho}) and the symmetry property of the inverse operator ${\bf R}$ it can be seen 
that the knowledge of the right hand side of (\ref{Rm1}) is enough to compute $\rho_k$
without solving explicitly the complicated linearized TBA equations (\ref{lin}) of the AdS/CFT.
Making use of (\ref{Rm1}) we get
\begin{equation}
\begin{split}
\rho_k
&=-\frac{1}{\pi}\sum_{m=1}^\infty\,\int_{-\infty}^\infty{\rm d}v\,
I_m(v)\,\partial_kL^\gamma_m(v)\\
&=\frac{1}{\pi}\sum_{m=1}^\infty\,\int_{-\infty}^\infty{\rm d}v\,
\int_{-\infty}^\infty{\rm d}u\,
s(v+i\gamma-u)\,Y^o_{m+1}(u)\,\partial_kL^\gamma_m(v)\\
&=\frac{1}{\pi}\sum_{m=1}^\infty\,\int_{-\infty}^\infty{\rm d}u\,
Y^o_{m+1}(u)\,\left(s\star\partial_kL_m\right)(u).
\end{split}
\end{equation}
This can be simplified further if we introduce the gauge transformed
T-system functions
\begin{equation}
\hat t_m(u)=\left\{\prod_kr_m(u-u_k)\right\}\,t_m(u),
\end{equation}
where
\begin{equation}
r_m(u)=\frac{1}{4}\,\frac{\gamma(2+m+iu)\,\gamma(2+m-iu)}
{\gamma(4+m+iu)\,\gamma(4+m-iu)}
\end{equation}
with $\gamma(z)=\Gamma(z/4)$. It is easy to check that in this gauge
we have
\begin{equation}
\hat t_m(u+i)\,\hat t_m(u-i)=1+y_m(u).
\end{equation}
Since $\hat t_m$ ($m\geq1$) has no roots in the physical strip we can write
\begin{equation}
\hat t_m(u)=\exp\left\{\left(s\star L_m\right)(u)\right\}
\end{equation}
and by taking the $\partial_k$ derivative we obtain
\begin{equation}
\partial_k\ln\hat t_m(u)=s\star\partial_kL_m(u).
\end{equation}
$\rho_k$ can now be written as
\begin{equation}
\rho_k=\frac{1}{\pi}\sum_{m=1}^\infty\,\int_{-\infty}^\infty{\rm d}u\,
Y^o_{m+1}(u)\,\partial_k\ln\hat t_m(u).
\end{equation}
Calculating the derivative we find
\begin{equation}
\partial_k\ln\hat t_m(u)=-{\cal F}_m(u-u_k)+\frac{2(u-u_k)}{m^2+(u-u_k)^2}
-\frac{u+u_k}{u^2-u_k^2+m(m+2)}.
\end{equation}
Putting everything together, we find the result
\begin{equation}
\begin{split}
\delta{\cal R}_k=
\frac{1}{\pi}\sum_{m=1}^\infty\,&\int_{-\infty}^\infty{\rm d}u\,
Y^o_m(u)\,\frac{u-u_k}{(m+1)^2+(u-u_k)^2}\\
&+\frac{1}{\pi}\sum_{m=1}^\infty\,\int_{-\infty}^\infty{\rm d}u\,
Y^o_{m+1}(u)\,\left\{\frac{u-u_k}{m^2+(u-u_k)^2}
-\frac{u_k}{u^2-u_k^2+m(m+2)}\right\}.
\end{split}
\end{equation}
This is precisely the same (up to a sign) as (\ref{Luscher}), the result obtained by using
the generalized L\"uscher formalism~\cite{BJ09}. Thus we have shown that up to 5-loop order
the TBA equations and the generalized L\"uscher formulae give the same result for the
anomalous dimension of the Konishi operator. 

\section*{Acknowledgments}

\'A. H. would like to thank Zolt\'an Bajnok
for useful discussions.
This work was supported by the Hungarian
Scientific Research Fund (OTKA) under the grant K 77400.

\appendix

\section{Existence and uniqueness of the inverse matrix}

The problem of finding the solution of the linearized TBA equations
(\ref{lin2}) is essentially equivalent to finding the inverse of the
infinite matrix (\ref{opmatrix}). In this appendix we show the existence and
uniqueness of this matrix inversion problem. Uniqueness, which is essentially
equivalent to the absence of zero modes, is important because this enables
us to calculate $\rho_k$ unambiguously from (\ref{intsol}). The infinite 
matrix (\ref{opmatrix}) can be written as
\begin{equation}
{\bf M}={\bf D}-{\bf P}\,s\star,
\label{opmatrix1}
\end{equation}
where ${\bf D}=<D_1^\gamma,D_2^\gamma,\dots>$ is diagonal and ${\bf P}$
is a constant tridiagonal matrix given by $P_{ij}=\delta_{i+1\,j}+
\delta_{i-1\,j}$. We can rewrite ${\bf M}$ as
\begin{equation}
{\bf M}=({\bf 1}-{\bf A}){\bf D}\,,\qquad{\rm where}\qquad 
{\bf A}={\bf P}s\star
{\bf D}^{-1}.
\end{equation}
The action of the operator ${\bf A}$ on a vector with components $f_i(x)$
can be written as 
\begin{equation}
({\bf A}f)_i(x)=\sum_jP_{ij}\int_{-\infty}^\infty{\rm d}y\,s(x-y)\,
d_j(y)f_j(y)\,,
\end{equation}
where $d_j(y)=1/D^\gamma_j(y)$. The crucial observation is that the absolute
value of this function is always smaller than its asymptotic value, $\Delta_j$:
\begin{equation}
\left\vert d_j(y)\right\vert<\Delta_j=\frac{j(j+2)}{(j+1)^2}\,
\qquad j=1,2,\dots,
\end{equation}
at least for small enough $\gamma$. For later use we now define the operator
${\bf B}$, which is obtained from ${\bf A}$ by replacing $d_j(y)$ with its 
asymptotic value:
\begin{equation}
({\bf B}f)_i(x)=\sum_jP_{ij}\int_{-\infty}^\infty{\rm d}y\,s(x-y)\,
\Delta_jf_j(y)\,.
\end{equation}
We also define analogously
\begin{equation}
{\bf M}_\infty=({\bf 1}-{\bf B}){\bf D}_\infty={\bf D}_\infty-{\bf P}s\star\,.
\end{equation}

The vectors of our linear space are given as infinite vectors
\begin{equation}
f\,\,\sim\,\,\left\{f_1(x),f_2(x),\dots\right\}\,,
\end{equation}
or, equivalently, in Fourier space as
\begin{equation}
f\,\,\sim\,\,\left\{\tilde f_1(\omega),\tilde f_2(\omega),\dots\right\}\,,
\end{equation}
where, as usual,
\begin{equation}
\tilde f(\omega)=\int_{-\infty}^\infty{\rm d}x\,{\rm e}^{ix\omega}\,f(x).
\end{equation}
We now equip our space with the hermitean scalar product
\begin{equation}
(g\vert f)=\sum_{i=1}^\infty\int_{\infty}^\infty{\rm d}x \,g^*_i(x)f_i(x)
=\frac{1}{2\pi}\,
\sum_{i=1}^\infty\int_{\infty}^\infty{\rm d}\omega \,\tilde g^*_i(\omega)
\tilde f_i(\omega)
\end{equation}
and the corresponding norm $\vert\vert f\vert\vert^2=(f\vert f)$. With this 
norm our vector space becomes a Hilbert space. We assume throughout this 
paper that both the vector variables $\xi$ and the source terms $j$ in 
equations of the form (\ref{eq}) belong to this Hilbert space. This is a
natural assumption since it is easy to see that both source terms 
(\ref{source1}) and (\ref{3.12}) and, more importantly, the vector on the
left hand side of (\ref{Rm1}) are elements of this Hilbert space.

For later purpose we note that the action of the operator ${\bf B}$ on the 
elements of the Hilbert space is simple in terms of the Fourier transformed
components. Using the notation ${\bf B}f=h$, we have
\begin{equation}
\tilde h_i(\omega)=\sum_jP_{ij}\tilde s(\omega)\Delta_j\tilde f_j(\omega)\,,
\end{equation}
where
\begin{equation}
\tilde s(\omega)=\frac{1}{2\cosh\omega}=\frac{1}{q+\frac{1}{q}}\,,\qquad
q={\rm e}^{\vert\omega\vert}\geq1.
\end{equation}

We now observe that
\begin{equation}
\begin{split}
\left\vert ({\bf A}f)_i(x)\right\vert&\leq\sum_jP_{ij}\int_{-\infty}^\infty
{\rm d}y\,s(x-y)\left\vert d_j(y)\right\vert\,\left\vert f_j(y)\right\vert\\
&<\sum_jP_{ij}\int_{-\infty}^\infty
{\rm d}y\,s(x-y)\Delta_j\,\hat f_j(y)=({\bf B}\hat f)_i(x)\,,
\end{split}
\end{equation}
where\footnote{All strict inequalities in this appendix are valid for
nonzero Hilbert space vectors $f,g$.}
\begin{equation}
\hat f_i(x)=\left\vert f_i(x)\right\vert\,,\qquad
(\hat f\vert\hat f)=(f\vert f).
\end{equation}
This inequality implies that ${\bf A}$ is \lq\lq smaller'' than ${\bf B}$,
in the sense that
\begin{equation}
\vert\vert{\bf A}f\vert\vert<\vert\vert{\bf B}\hat f\vert\vert\qquad{\rm and}
\qquad
\left\vert (g\vert{\bf A}f)\right\vert<(\hat g\vert{\bf B}\hat f).
\end{equation}
On the other hand, ${\bf B}$ is smaller than unity, in the following sense.
We first write
\begin{equation}
(g\vert {\bf B}f)=\frac{1}{2\pi}\int_{-\infty}^\infty{\rm d}\omega
\tilde s(\omega)\sum_{j=1}^\infty\Delta_j\tilde f_j(\omega)\left\{
\tilde g^*_{j+1}(\omega)+\tilde g^*_{j-1}(\omega)\right\}
\end{equation}
and after using the simple inequality $2\vert ab\vert\leq\vert a\vert^2+
\vert b\vert^2$ we have
\begin{equation}
\left\vert(g\vert {\bf B}f)\right\vert<
\frac{1}{2\pi}\int_{-\infty}^\infty{\rm d}\omega
\tilde s(\omega)\sum_{j=1}^\infty\left\{
\left\vert \tilde f^*_j(\omega)\right\vert^2+
\left\vert\tilde g^*_j(\omega)\right\vert^2\right\}<
\frac{1}{2}\,\vert\vert f\vert\vert^2+\frac{1}{2}\,\vert\vert g\vert\vert^2\,.
\end{equation}
Thus the norm of ${\bf B}$ is not exceeding unity since from the above
inequality it follows that
\begin{equation}
\vert(f\vert{\bf B}f)\vert<(f\vert f)
\label{un1}
\end{equation}
and similarly
\begin{equation}
\vert(f\vert{\bf A}f)\vert<(\hat f\vert{\bf B}\hat f)<
(\hat f\vert\hat f)=(f\vert f)\,.
\label{un2}
\end{equation}
The inequalities (\ref{un1}) and (\ref{un2}) imply {\it uniqueness}
of the inverse of the operators ${\bf 1}-{\bf B}$ and ${\bf 1}-{\bf A}$
since by multiplying the equations
\begin{equation}
f={\bf B}f\,,\qquad\quad f={\bf A}f
\end{equation}
by $f$ we arrive at a contradiction.

More precisely, the solution of $({\bf 1}-{\bf A})f=0$ as an infinite
component vector $\{f_1(x),f_2(x),\dots\}$ may formally exist, but the
above considerations show that $f$ cannot be a vector of the Hilbert space.
The analogous ${\bf M}_\infty \xi=0$ equation can be explicitly solved in
Fourier space. This corresponds to the recursion relation
\begin{equation}
\left(q+\frac{1}{q}\right)\,\frac{(k+1)^2}{k(k+2)}\,\tilde\xi_k=
\tilde\xi_{k+1}+\tilde\xi_{k-1}\,,\qquad k=1,2,\dots
\label{rec}
\end{equation}
with the boundary condition $\tilde\xi_0=0$. The formal solution is easily
found:
\begin{equation}
\tilde\xi_k(\omega)=C_1(\omega)a(k)\,,\qquad
a(k)=\frac{k}{k+1}\left(q^{k+2}-q^{-k-2}\right)-
\frac{k+2}{k+1}\left(q^k-q^{-k}\right)\,.
\end{equation}
Here $C_1(\omega)$ is an arbitrary ($\omega$-dependent) normalization
constant. 
Of course, this $\xi$ cannot be an element of the Hilbert space, since its 
components are exploding in $k$. This shows why the Hilbert 
space requirement is natural: linearization only makes sense as long as
the linearized variable remains small.

The general solution of the recursion relation
(\ref{rec}) is
\begin{equation}
\tilde\xi_k(\omega)=C_1(\omega)a(k)+
C_2(\omega)b(k)\,,\qquad
b(k)=\frac{k+2}{k+1}\,q^{-k}-\frac{k}{k+1}\,q^{-k-2}\,,
\end{equation}
where $C_2(\omega)$ is a second normalization constant. Using the building 
blocks $a(k)$ and $b(k)$ the inverse of ${\bf M}_\infty$ in Fourier space 
can be written as
\begin{equation}
\tilde R_{\infty\,kl}(\omega)=\left\{
\begin{aligned}
\lambda(\omega)\,a(k)b(l)\qquad k\leq l\,,\\
\lambda(\omega)\,a(l)b(k)\qquad k\geq l\,,
\end{aligned}
\right. 
\end{equation}
where
\begin{equation}
\lambda(\omega)=\frac{\cosh\omega}{4\sinh^3\vert\omega\vert}.
\end{equation}
One can show that
\begin{equation}
{\bf R}_\infty={\bf D}_\infty^{-1}({\bf 1}-{\bf B})^{-1}={\bf D}_\infty^{-1}
{\bf b}\,,
\end{equation}
where ${\bf b}$ is the sum of the Neumann series
\begin{equation}
{\bf b}=({\bf 1}-{\bf B})^{-1}={\bf 1}+{\bf B}+{\bf B}^2+\dots
\end{equation}
Since ${\bf A}$ is \lq\lq smaller'' than ${\bf B}$, the inverse of
${\bf 1}-{\bf A}$ also exists in the form of the Neumann series
\begin{equation}
{\bf a}=({\bf 1}-{\bf A})^{-1}={\bf 1}+{\bf A}+{\bf A}^2+\dots
\end{equation}
since it can be shown easily that
\begin{equation}
\vert\vert({\bf 1}+{\bf A}+{\bf A}^2+\dots+{\bf A}^k)f\vert\vert<
\vert\vert({\bf 1}+{\bf B}+{\bf B}^2+\dots+{\bf B}^k)\hat f\vert\vert<
\vert\vert{\bf b}\hat f\vert\vert\,.
\end{equation}
It is also evident that the inverse of ${\bf M}$, which can be written as
\begin{equation}
{\bf R}={\bf D}^{-1}+{\bf D}^{-1}{\bf P}s\star{\bf D}^{-1}+
{\bf D}^{-1}{\bf P}s\star{\bf D}^{-1}{\bf P}s\star{\bf D}^{-1}+\dots
\end{equation}
is manifestly symmetric.


\end{document}